# *p* x *n*-Type Transverse Thermoelectrics in a Type-II Weyl Semimetal TaIrTe$_4$

Josh Mutch[1], Wei-Chih Chen[2], Cheng-Yi Huang[3], Paul Malinowski[1], Cheng-Chien Chen[2], Jiun-Haw Chu[1,*]

[1] Department of Physics, University of Washington, Seattle, WA 98105

[2] Department of Physics, University of Alabama at Birmingham, Birmingham, Alabama 35294, USA

[3] Department of Physics, Northeastern University, Boston, Massachusetts 02115, USA

*jhchu@uw.edu

**Abstract**

*p* x *n*-type materials refer to materials with a *p*-type Seebeck coefficient in one direction and a *n*-type coefficient in the orthogonal direction. This type of materials allows for a transverse thermoelectric response, which is highly desirable for energy applications. Here, we report the observation of *p* x *n*-type behavior in TaIrTe$_4$, a type-II Weyl semimetal, with an in-plane thermopower anisotropy $S_{xx} - S_{yy}$ reaches a maximum value $40 \mu V/K$ at 200K. Intriguingly. we found that such a *p* x *n*-type behavior is absent in the similar compound NbIrTe$_4$. The presence and absence of *p* x *n*-type behavior in these two materials are consistent with density functional theory calculations, which further predict that the thermopower anisotropy in both compounds can be enhanced up to 130 $\mu V/K$ by electron doping. Such a strong thermopower anisotropy originates from the presence of both *p*-type and *n*-type carriers, each with high mobility in one direction. These results suggest that although type-II Weyl semimetal phase does not guarantee the existence of *p* x *n*-type behavior, its unique band structure provides the ingredient to engineer and optimize this phenomenon.

**Introduction**

Transverse thermoelectric effect refers to the conversion of a temperature gradient into a charge current in the orthogonal direction. In contrast to longitudinal thermoelectrics, devices based on transverse thermoelectrics have simpler structure and higher efficiency[1]. In general, transverse thermoelectric effect requires an off-diagonal term in the thermoelectric tensor that is only allowed by breaking time reversal or mirror symmetry. The former has motivated the search for magnetic materials with large anomalous Nernst effect[2-6], whereas the latter can be achieved by materials with large anisotropic Seebeck coefficient, such as *p* x *n*-type materials [7,8]. In *p* x *n*-type materials, the Seebeck coefficient is *p*-type along one crystalline axis and *n*-type along the orthogonal axis. By choosing a proper direction in between the two principal axes, an off-diagonal term $(S_{xx} - S_{yy})\cos\theta\sin\theta$ naturally emerges in the rotated frame, where θ is the angle of rotation. Despite the active searching in the past decades[9], only a handful of *p* x *n*-type materials have been identified [10]. Even fewer show a peak thermopower anisotropy greater than 5 $\mu V/K$ or that persists beyond low temperatures[11-17].

In this study, we report the observation of *p* x *n* type behavior in single crystals of TaIrTe$_4$, a type-II Weyl semimetal with a van der Waals layered crystal structures[18-20]. The Seebeck coefficient reaches -15 $\mu V/K$ along the crystal *a*-axis and 25 $\mu V/K$ along the crystal *b*-axis. Interestingly, this behavior is absent in NbIrTe$_4$, which is also a type-II Weyl semimetal[21,22]. Density functional theory (DFT) band structure calculations are consistent with these experimental results and indicate that the *p* x *n*-type behavior is sensitive to the position of the Fermi level. The DFT calculations further predict that tuning the Fermi level of this class of materials via electron doping enhances the transverse thermoelectric coefficient by threefold.

**Methods**

Bulk single crystals of TaIrTe$_4$ and NbIrTe$_4$ were grown out of Te flux following procedures similar to previous reports [19]. Ta or Nb powder, Ir powder, and Te shot were mixed in an atomic ratio of 1:1:20 with a total mass of starting material ~3.5 g and loaded into Canfield crucible sets[23] which were then sealed in quartz tubes under a low-pressure argon environment. The growths were placed into box furnaces and heated to 1000 °C over 12 hours, held at 1000 °C for 24 hours, and then slowly cooled to 700 °C over 150 hours. The growths were then decanted via centrifuge to separate the excess Te flux from the resulting single crystals. For both compounds, the yield resulted in single crystals with shiny facets and ribbon or needle morphology.

Thermoelectric measurements were performed in a PPMS Dynacool using a setup that was first introduced by Mun et al[24]. To eliminate the offset error resulting from thermal emfs, we employed a "heater switching" method. By alternating heating each end of the samples, unwanted thermal emfs can be eliminated by extracting the difference of the sample voltage measured in each configuration. Details of instrumentation can be found in ref[25].

DFT calculations were performed using the Vienna Ab initio Simulation Package (VASP) [26,27], which adopts a pseudopotential method and a plane-wave basis. We first utilized the non-local van der Waals density functional, optB88-vdW [28], to account for London dispersion interactions during structure relaxation. The self-consistent electronic structures were calculated using the Perdew-Burke-Ernzerhof functional (PBE-GGA) [29] with spin-orbit interaction. The structure relaxation was performed until the force on each atom is less than $10^{-3}$ eV/Å, and the convergence criterion for self-consistent field is set to $10^{-6}$ eV/unit-cell. A kinetic energy cutoff of 400 eV and a Monkhorst-Pack k-grid [30] of 17x7x7 were adopted after careful convergence tests. With the DFT electronic structures as inputs, Wannier90 [31] was then employed to obtain the corresponding tight-binding models. For transport properties, we used Boltzmann transport equation implemented in the BoltzWann code [32] with a dense k-grid of 200x56x56, which is sufficient for temperatures above 150 K based on our convergence tests.

**Results and Discussions**

TaIrTe$_4$ and NbIrTe$_4$ both crystallize in a noncentrosymmetric orthorhombic structure of space group *Pmn2$_1$* [19,22], as shown in Fig. 1(a). Tellurium octahedra structures form in the *a/b* plane, with alternating tantalum and iridium atoms contained within the octahedra structures. These layers van der Waals couple in the out of plane *c* lattice direction. We use the convention *x:y:z* to refer to the *a:b:c* lattice directions of the crystal. Figure 1(b) shows the main result of our study: The Seebeck coefficient of TaIrTe$_4$ changes sign along different axes, measuring *n*-type in the *x* direction and *p*-type in the *y* direction. This directional dependence persists from 20K to the maximum temperature we measured at 300K and attains a maximum difference of ($S_{yy}$ - $S_{xx}$) ~ 40 $\mu V/K$ at 200K. Below 20K, small magnitude *p*-type behavior is measured in both directions. We also performed resistivity measurements along the in-plane axes, as shown in Fig. 1(c). TaIrTe$_4$ displays a strong in-plane resistivity anisotropy, with $\rho_{yy}$ three to six times greater than $\rho_{xx}$ over the measured temperature range.

The in-plane thermopower and resistivity for NbIrTe$_4$ are shown in Fig. 2. Unlike TaIrTe$_4$, the sign of the Seebeck coefficient for NbIrTe$_4$ is the same in both the *x* and the *y* directions, displaying *n*-type behavior below 90K and *p*-type behavior above 130 K. Only for a very narrow temperature range (90 K < T < 130 K) does NbIrTe$_4$ display *p* x *n* behavior, with a much smaller anisotropy ($S_{yy}$ - $S_{xx}$) ~ 2 $\mu V/K$. Interestingly, the resistivity anisotropy is much greater in NbIrTe$_4$ compared to TaIrTe$_4$, with the ratio $\rho_{yy}/\rho_{xx} > 11$ over the entire temperature range measured.

The most common mechanism responsible for *p* x *n*-type behavior is the multiband effect, where *p*-type bands and *n*-type bands dominate conduction in different directions. In a two-band model of one *p*-type band and one *n*-type band, the total Seebeck tensor can be expressed as:

$$S = (\sigma_p + \sigma_n)^{-1}(\sigma_p S_p + \sigma_n S_n) \qquad 1$$

Indeed, DFT calculations suggested that TaIrTe$_4$ and NbIrTe$_4$ host two to three closed hole pockets and two open electron pockets [19,22], which have been corroborated by the quantum oscillations and angle resolved photoemission experiments [20,21]. In addition to the coexistence of electrons and holes, the orthorhombic crystalline symmetry of TaIrTe$_4$ and NbIrTe$_4$ indicates that these carriers are likely anisotropic, making them an ideal platform to search for *p x n* type behavior [33-36].

Clearly the coexistence of electron and hole pockets in an orthorhombic crystal lattice does not guarantee the existence of *p* x *n* type behavior since this behavior is not always present in NbIrTe$_4$. To maximize *p* x *n* type behavior, the anisotropy of electron and hole pockets should be opposite. The absence of *p* x *n* type behavior combined with a stronger resistivity anisotropy in NbIrTe$_4$ compared to TaIrTe$_4$ suggests that the *p*-type carriers effectively dominate thermopower and conduction in both directions, whereas in TaIrTe$_4$ each carrier type dominates conduction in different direction. This is corroborated by the higher resistivity anisotropy in NbIrTe$_4$, where the total conductivity tensor is dominated by the highly anisotropic *p*-type carriers. In TaIrTe$_4$, the strong competition between the *p*-type and *n*-type carriers leads to a reduced resistivity anisotropy since the high mobility axis of each carrier type is rotated with respect to one another.

To gain more insight, we performed DFT calculations to study the electronic and transport properties. Figure 3(a) and (d) display the electronic band structures respectively for TaIrTe$_4$ and NbIrTe$_4$. Overall, their band structures resemble each other due to the similarity in the crystal structures and atomic compositions. However, their transport and thermoelectric properties can sensitively depend on the details of the band structure near the Fermi level. Figures 3(b) and 4(c) show the Seebeck coefficients of TaIrTe$_4$ as a function of the chemical potential along the *x*- and *y*-axis, respectively. At the Fermi level, the $S_{xx}/S_{yy}$ has a value of ~-30/+25 $\mu V/K$ at 200 K, which yields a thermopower anisotropy of 55 $\mu V/K$ in agreement with the experimental result of a prominent *p* x *n* behavior in TaIrTe$_4$. The calculations of $S_{xx}$ and $S_{yy}$ for NbIrTe$_4$ are shown in Fig. 4(e) and 4(f), respectively. At 200 K, the theoretical $S_{xx}/S_{yy}$ at the Fermi level has a value of ~-1/-13 $\mu V/K$, indicating a much weaker *p* x *n* behavior in NbIrTe$_4$ also consistent qualitatively with the experiment. A possible source of theory-experiment discrepancy could come from the uncertainty in determining the chemical potential.

Intriguingly, the DFT calculations also indicate that increasing the Fermi level by 200 meV for both TaIrTe$_4$ and NbIrTe$_4$ could create an extraordinary thermopower anisotropy, increasing the transverse signal by an additional factor of ~3. Given the agreement between the calculations and experiments on the stoichiometric TaIrTe$_4$ and NbIrTe$_4$, we believe that the electronic doping is a very promising direction to further optimize the *p* x *n* thermoelectricity in this class of materials.

We close by commenting on the topological nature of TaIrTe$_4$ and NbIrTe$_4$ and its relation to *p* x *n* type behavior. Both TaIrTe$_4$ and NbIrTe$_4$ are type-II Weyl semimetals. In a type-II Weyl semimetal, the Weyl crossing is tilted such that the Fermi velocity changes sign at certain direction of the momentum. Consequently, electron and hole pockets always coexist when the Fermi level is close to type-II Weyl points. This does not explain the coexistence of electron and hole pockets in TaIrTe$_4$ and NbIrTe$_4$, because the Weyl points in both materials are 70-100 meV away from the Fermi level, well above the Lifshitz energy. Nevertheless, we notice that there is a substantial overlap between conduction bands and valence bands in these two materials as can be seen from the DFT calculations. Such an overlap is necessary to generate the type-II Weyl crossing and is also likely to form the coexisting electron and hole pockets for realizing the *p* x *n*-type behavior.

**Conclusion**

Despite the exciting engineering advantage offered by transverse thermoelectric devices, the search for highly efficient transverse thermoelectric materials is a present challenge. It was recently shown that a 4-6 $\mu V/K$ Nernst effect in an iron-based compound could achieve large power generation capacity in micro-scale devices [4]. In this study, we have measured the thermopower and resistivity anisotropies of TaIrTe$_4$ and NbIrTe$_4$. TaIrTe$_4$ displays strong thermopower anisotropy and *p* x *n* behavior, while the thermopower of NbIrTe$_4$ is much more isotropic. Both the Ta and Nb compound display significant thermopower anisotropy if the Fermi energy is tuned 200 meV higher, as predicted by DFT calculations. The presence of both *p*-type and *n*-type carriers revealed by DFT calculations suggests that the multi-band effect creates the *p* x *n* behavior in these materials. While any applications with these specific materials are limited due to the low earth abundance of iridium, the observation of *p* x *n* behavior opens the door for a search of more materials within the type-II Weyl semimetal family with potential applications in transverse thermoelectrics.

**Acknowledgments**

The research is supported by the Air Force Office of Scientific Research (AFOSR) under Award No. FA-2386-21-1-4060. Materials synthesis at UW was supported as part of Programmable Quantum Materials, an Energy Frontier Research Center funded by the U.S. Department of Energy (DOE), Office of Science, Basic Energy Sciences (BES), under award DE-SC0019443 and the Gordon and Betty Moore Foundation's EPiQS Initiative, Grant GBMF6759 to JHC. The calculations were performed on the Frontera computing system at the Texas Advanced Computing Center. Frontera is made possible by the National Science Foundation (NSF) Award No. OAC-1818253. J.-H.C. also acknowledges the support from the State of Washington funded Clean Energy Institute

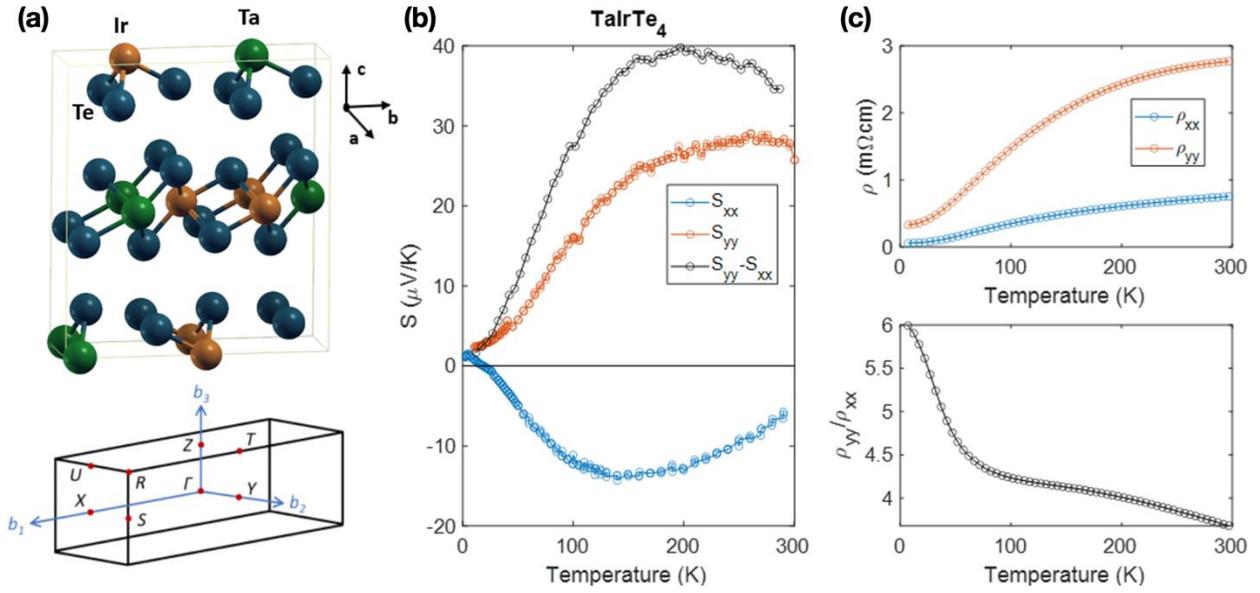

FIG. 1: (a) Crystal structure [top panel] and Brillouin zone [bottom panel] of TaIrTe$_4$. Tellurium atoms are shown in blue, iridium in orange, and tantalum in green. (b) Thermopower measurements for TaIrTe$_4$ along the $x$ direction (blue) and $y$ direction (red), with the difference ($S_{yy}$ - $S_{xx}$) shown in black. (c) The resistivity measured along the $x$ (blue) and $y$ (red) direction. In the lower panel, the resistivity ratio $\rho_{yy}/\rho_{xx}$ is plotted as a function of temperature.

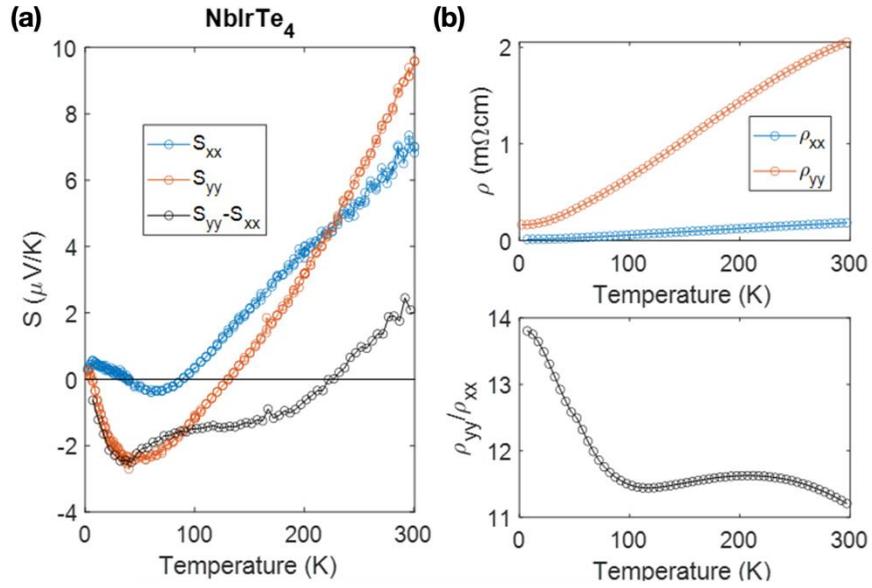

FIG. 2: (a) Thermopower measurements for NbIrTe$_4$ along the $x$ direction (blue) and $y$ direction (red), with the difference ($S_{yy}$ - $S_{xx}$) shown in black. (b) The resistivity measured along the $x$ (blue) and $y$ (red) direction. In the lower panel, the resistivity ratio $\rho_{yy}/\rho_{xx}$ is plotted as a function of temperature.

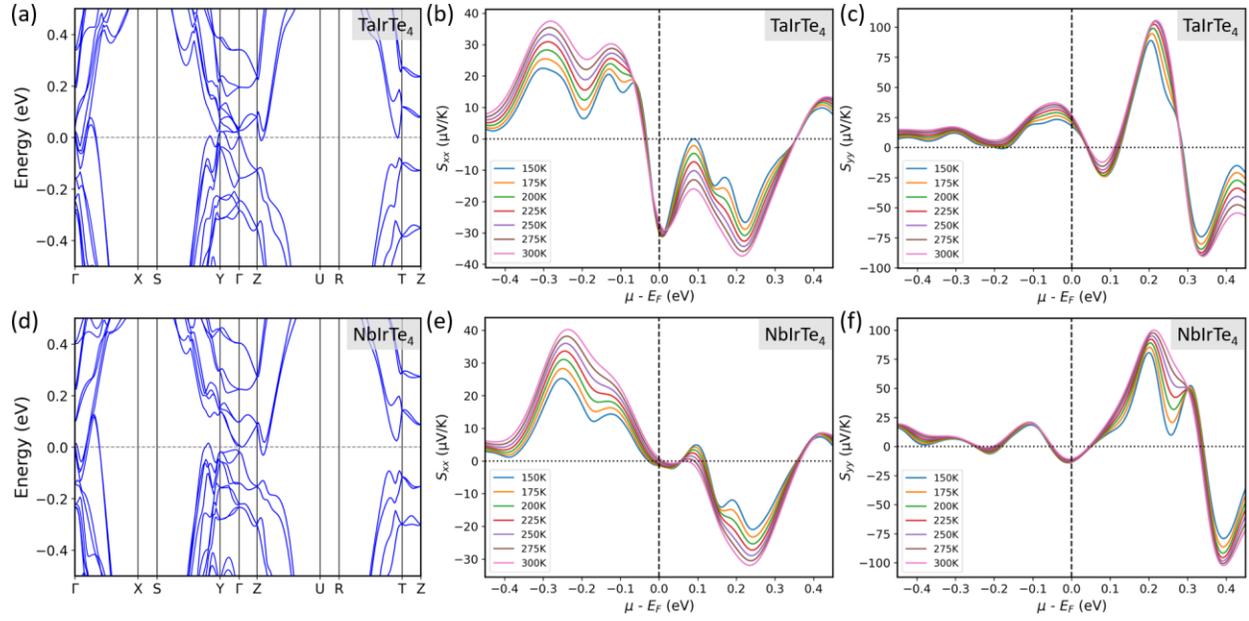

FIG. 3: DFT electronic band structures along the high symmetry points in the Brillouin zone for (a) TaIrTe$_4$ and (d) NbIrTe$_4$. DFT Seebeck coefficients of TaIrTe$_4$ along the (b) *x*- and (c) *y*-axis, and those of NbIrTe$_4$ along the (e) *x*- and (f) *y*-axis over a temperature range of 150 K- 300 K, as denoted in the legend.